# Callisto III e la Cometa di Halley: la ricerca di Johan Stein SJ tra leggenda e storia

## Costantino Sigismondi

Observatorio Nacional, Rio de Janeiro, ICRA e Ateneo Pontificio Regina Apostolorum, Roma

**Abstract:** The Dutch Jesuit astronomer Johan Stein published in 1909 a text about the legend on a papal bull of Pope Callixtus III against the comet of 1456. This comet happened to be the Halley's one. The original documents, either chronicles and observations, and coeval testimonies are deeply investigated by Stein and the falsity of that legend is clearly demonstrated.
In the occasion of the 200 years of the restoration of the *Societas Jesu* made in 1814 by Pope Pius VII an Italian edition of the full text of Johan Stein is here presented.

## Introduzione

Il corposo articolo del padre olandese Johan Stein SJ (1871-1951),[1] scritto quando era assistente del direttore della Specola Vaticana padre Johann Georg Hagen, mira a sradicare la favola della scomunica papale di Callisto III del 1456 contro la cometa,[2] che quasi tre secoli dopo avrebbe portato il nome di Halley.
Piú tardi, nel 1930, anche padre Stein divenne il direttore della Specola, contribuí alla sua modernizzazione e trasferimento in Castelgandolfo (1933) e ne scrisse una storia.[3]
Nello stesso articolo, apparso sul bollettino della Specola, *Ricerche Astronomiche*, lo Stein riporta un abstract in Italiano del suo studio del 1909 sulla alla bolla di Callisto III[4] condotto insieme all'archivista vaticano Ranuzzi sui documenti originali, tratteggiando come l'improbabile leggenda sarebbe una conseguenza che la bolla fu promulgata poco dopo l'apparizione della cometa e dei relativi pronostici astrologici.
Questo lavoro, originalmente scritto in francese, oggi viene riproposto per la prima volta integralmente in lingua Italiana per i tipi dell'Ateneo Pontificio Regina Apostolorum, segnatamente nella collana Scienza e Fede diretta dal P. Rafael Pascual,
e vede la luce nell'anno bicentenario della restaurazione della Compagnia di Gesú avvenuta nel 1814 sotto il papa Pio VII Chiaramonti, durante il pontificato del primo papa Gesuita, Francesco, Jorge Mario Bergoglio.

## Scienza e Informazione durante il passaggio del 1910

L'articolo dell'astronomo vaticano usciva poco prima del passaggio "spaventoso" della cometa di Halley del 1910, dove addirittura la Terra doveva passare dentro la coda della cometa il 19 di maggio. Qualcuno pensando ai miasmi della cometa pensó bene di fornire i primi automobilisti di una bottiglia di acqua purissima.

---

[1] http://en.wikipedia.org/wiki/Johan_Stein
[2] http://en.wikipedia.org/wiki/Pope_Callixtus_III#The_.22bull_against_the_comet.22
[3] JOHAN STEIN, Cinquanta anni di attività della Specola Vaticana (1891-1941), Ricerche Astronomiche 1, 135 (1942).
[4] *ibidem*, p. 140.

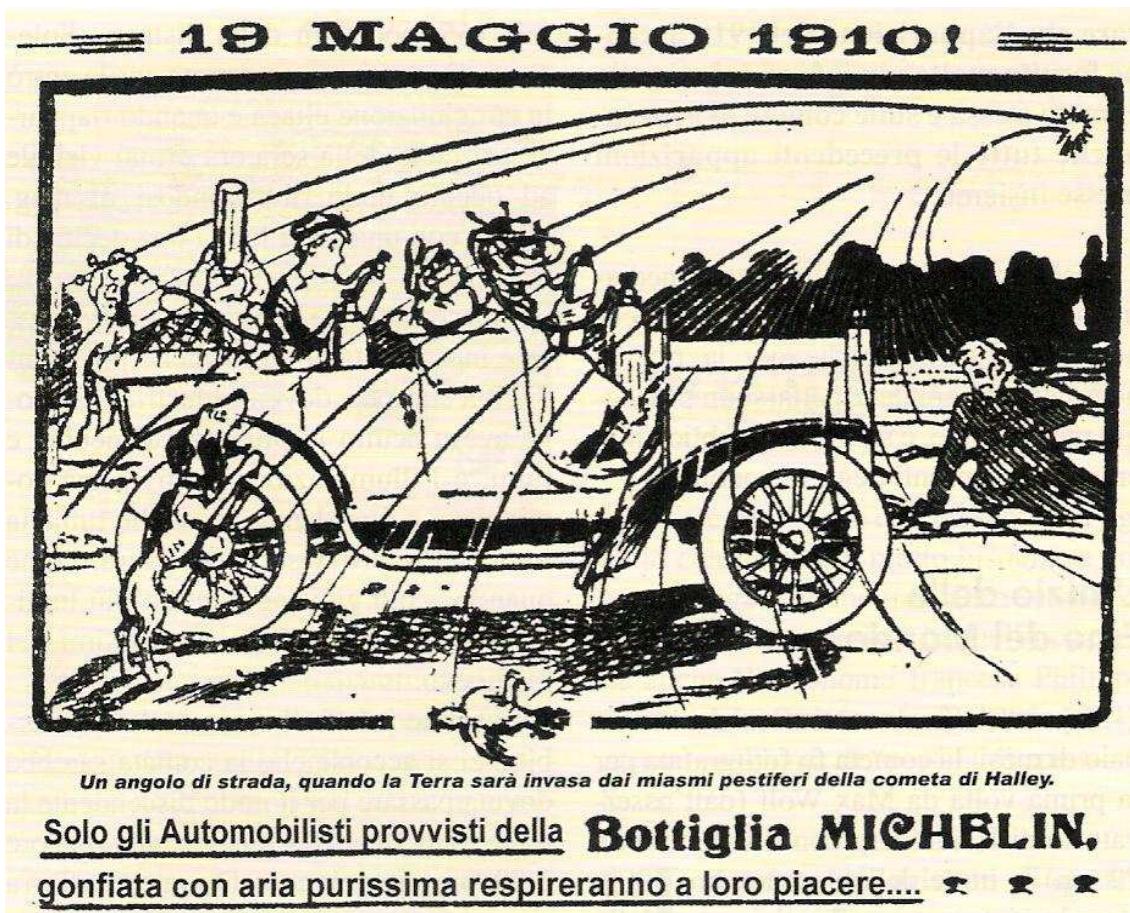

**Fig. 1** Vignetta sulle conseguenze del passaggio della Terra nella coda della Cometa di Halley nel 1910.

Annunci di sventure e apocatastasi non mancarono sia pure in pieno positivismo...
ma la differenza tra scienza e fantascienza non è cosí netta come si potrebbe pensare, e la panspermia legata al nome di sir Fred Hoyle ripeté in chiave tecnico scientifica il tema della vignetta della bottiglia Michelin[5] alcuni decenni dopo.
Leggiamo la prima traduzione Italiana, annotata, oltre 100 anni dopo la sua stesura originale, dopo che la cometa di Halley ritornò al perielio nel 1985-86, verificando anche le previsioni razionalistiche pubblicate dal Babinet nel 1853 riguardo alla popolosità dell'America Settentrionale, ma non quelle su una megalopoli panamense.[6]

**Metodologia nello studio di J. Stein SJ**

Il rigore storico delle ricerche dello Stein è unito anche ad una delicatezza nei confronti dei protagonisti che non porta mai l'autore a 'scomunicare' nessuno, nemmeno Pierre Simon de Laplace dalla cui somma penna prese piede la favola della scomunica papale alla Cometa.
Anzi lo stesso Stein, con una ricerca bibliografica sul giornale *La Quotidienne*, di mercoledì 7 marzo 1827 (n. 66), p. 2, ha cura di far notare come Laplace sia morto in pace con Dio, con il conforto dei sacramenti il 5 marzo 1827 nella sua casa di rue du Bac a Parigi,[7] a conclusione di una vita sempre intellettualmente onesta.
Questo metodo consente di non creare inutili testa a testa tra Scienza e Fede, e si rivela perciò molto fecondo anche per il mondo attuale.

---

[5] FRED HOYLE, CHANDRA WICKRAMASINGHE AND JOHN WATSON, *Viruses from Space and Related Matters*, University College Cardiff Press (1986).
[6] Questa ed altre curiosità sono riportate dall'articolo del Padre Stein nel libro in oggetto.

**Il passaggio del 1985-86 della cometa di Halley nel lavoro di Paolo Maffei**

Col passaggio del 1985-86 una nuova ondata editoriale a livello internazionale ha eclissato tutti i lavori del 1909-10, sia pure mettendoli come referenze.
La monografia di Paolo Maffei dei 1984 per i tipi di Mondadori,[8] nel panorama delle pubblicazioni in Italiano, sia per l'autorità di Maffei, sia per le referenze ed i commenti, il lavoro più importante tra tutti quelli pubblicati in quegli anni.
Maffei reintrodusse per il grande pubblico i documenti degli astronomi Cinesi, Coreani e Giapponesi per documentare tutti i passaggi della cometa di Halley storicamente accessibili.
Le mappe stellari, anche quelle vergate da Paolo dal Pozzo Toscanelli in occasione del passaggio del 1456, complementano quella preziosa edizione.
Maffei attribuisce a Giotto, pittore della cappella degli Scrovegni poco dopo il passaggio della cometa di Halley del 1304, l'invenzione della stella con la coda per rappresentare la stella dei Magi, oggi in tutti i presepi.
Giotto, primo tra i pittori realisti, si sarebbe ispirato proprio alla cometa di Halley, o ad una passata a qualche mese di distanza, per rappresentare la natività.

Sulla questione di Callisto III, Maffei non va oltre l'aneddotica normale, citando le preghiere "contra turcos cometamque", ma poteva dedicare lo stesso spazio di Stein, poiché il suo obbiettivo era documentare oltre 2000 anni di passaggi al perielio della Halley al fine di ricostruirne tutti gli elementi orbitali ed il contributo delle forze non gravitazionali nelle varie orbite.
Fa fede il fatto che il testo di Maffei abbia in appendice le effemeridi della cometa del 1985-86, usate da chi scrive nelle osservazioni fatte secondo le indicazioni[9] della campagna "International Halley Watch".[10]

L'opera dello Stein aiuta a convincere il lettore che mai la Chiesa Cattolica poteva pensare di scomunicare un fenomeno celeste, nonostante che la sua natura astronomica fosse stata provata definitivamente solo da Tycho Brahe dopo la grande cometa del 1577.[11]

---

[7] Al 140 di Rue du Bac, nella cappella delle suore di s. Vincenzo de Paoli e s. Luisa de Marillac, la Vergine Maria apparve a santa Caterina Labouré il 18 luglio 1830, oggi a ricordo c'è un santuario e la devozione alla medaglia miracolosa.

[8] PAOLO MAFFEI, La cometa di Halley, Mondadori, Milano (1984),

[9] STEPHEN J. EDBERG, *International Halley Watch Amateur Observers' Manual for Scientific Comet Studies*, Enslow Publisher and Sky Publishing Corp. (1983).

[10] Risultati della campagna osservativa sul sito
http://pdssbn.astro.umd.edu/data_sb/missions/ihw/index.shtml

[11] J. R. CHRISTIANSON AND TYCHO BRAHE, *Tycho Brahe's German Treatise on the Comet of 1577: A Study in Science and Politics,* Isis, **70**, 110-140 (1979).

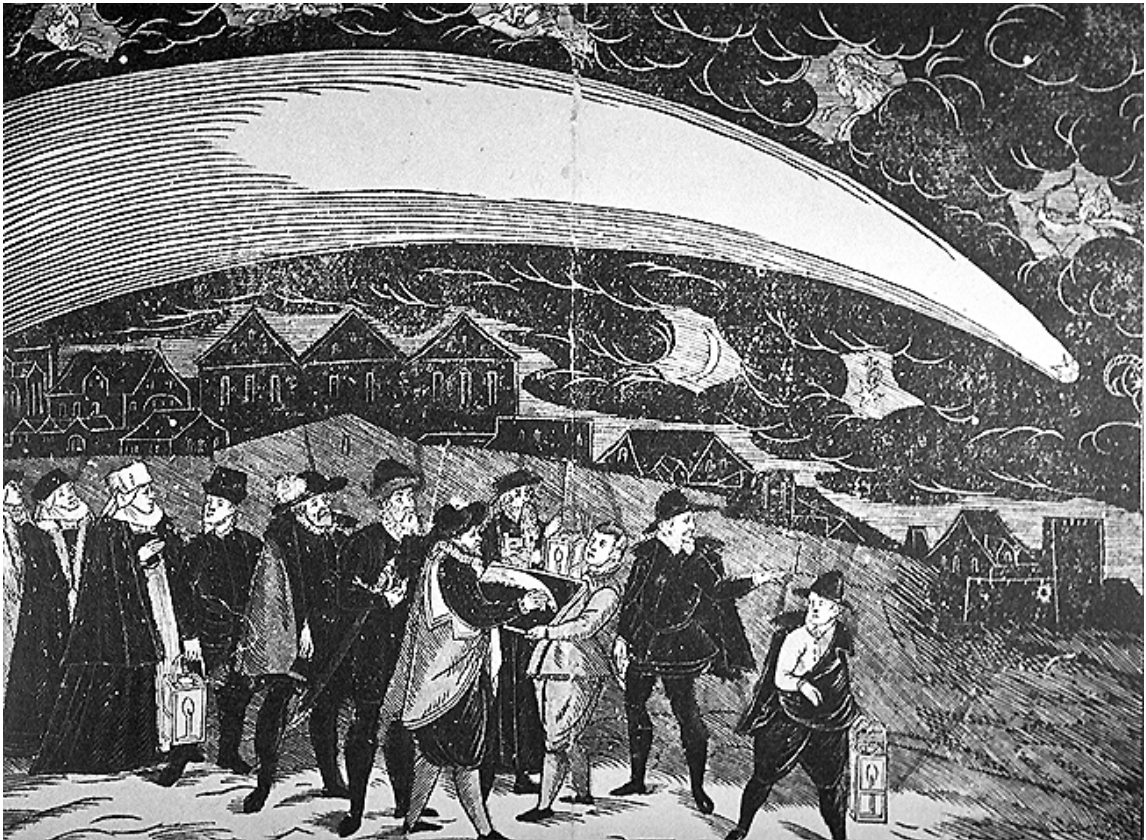
**Fig. 2** La grande Cometa del 1577 vista da Praga.

**Conclusioni**

È raccomandabile la lettura di questo libro di Stein a quanti vogliano imparare da un maestro il metodo della ricerca storica nel dialogo tra Scienza e Fede: una ricerca fatta con precisione e delicatezza nel giudizio.

Dalla citazione dell'immortale Newton[12] sulla natura delle stelle *Novae*, alimentate dalle comete, a quella di Beda il Venerabile[13] sugli influssi delle comete sulle passioni umane, il testo contiene anche un eccellente panorama sulle idee riguardo alle comete e ai loro influssi apparse nel corso della storia.

Pio II, Enea Silvio Piccolomini e Paolo II, Pietro Barbo sono contemporanei di Callisto III Borgia e co-protagonisti di queste vicende.

Un altro papa, Alessandro VII Chigi, appare in una lettera di Athanasius Kircher SJ in parzialmente ripubblicata dallo Stein, in merito alla sua posizione sulle comete e l'astrologia. E' interessante che il Kircher, Gesuita dal sapere enciclopedico, dichiari di aver trattato della cometa del 1664 *il meno possibile tra tanti cuochi indaffarati a preparare questo brodetto!*

Talvolta nel dialogo tra scienza e fede, anche il silenzio puo' essere opportuno.

---

[12] Isaac Newton, *Philosophiae naturalis principia mathematica* (ed. 1713), prop. 42, probl. 22 in fine.
[13] S. Beda il Venerabile (677-735), *De natura rerum*, c. 24; Migne, P. L./PL, 90, col. 243-244.

**Referenze**